\documentclass[twoside]{article}
\usepackage{fleqn,valencia}

\usepackage{epsfig}

\newcommand{\AmS}{{\protect\the\textfont2
  A\kern-.1667em\lower.5ex\hbox{M}\kern-.125emS}}

\title{ Dalitz plot analysis and branching fraction measurement of 
$D^+$ and $D^+_s\to \pi^-\pi^+\pi^+$ decays \thanks{  Talk presented 
	in  4th International Conference Hyperons, Charm
	        and Beauty Hadrons in Valencia June 2000.}}

\author{Ignacio Bediaga\address[cbpf]{Centro Brasileiro de Pesquisas F\'\i sicas, \\ 
        Rua Xavier Sigaud 150, 22290, Rio de Janeiro, Brazil}%
	\thanks{ Representing the Fermilab E791 Collaboration.}}

\begin{document}

\begin{abstract}
 Fermilab fixed target experiment E791 obtained a sample 
 of $ 1172 \pm 61 $ events of $ D^+ \to \pi^- \pi^+ \pi^+ $ 
and   848  $\pm 44$ events of $D_s^+ \to \pi^- \pi^+ \pi^+$
 decays. We find respectively 
 $ B (D^+ \to \pi^- \pi^+ \pi^+ ) / B ( D^+ 
 \to K^- \pi^+ \pi^+ ) = 0.0311 \pm 0.0018 ^
 {+0.0016}_{-0.0026} $ and  $B(D_s^+ \to \pi^+ \pi^- \pi^+)  
 / B(D_s^+ \to \phi \pi^+)  = 0.245 \pm 
 0.028^{+0.019}_{-0.012} $. Using a coherent amplitude 
 analysis to fit the Dalitz plot of the $ D^+ \to \pi^- \pi^+ 
 \pi^+ $ decay, we find strong evidence  for a  scalar resonance 
 of mass $478^{+24}_{-23} 
 \pm 17 $ MeV/$c^2$ and width $324^{+42}_{-40} \pm 21 $ 
 MeV/$c^2$, compatible with what is expected for the 
 isoscalar meson  $\sigma$ . The $ D^+ \to \sigma(500) 
 \pi^+ $ fraction accounts for approximately half of all 
 three-charged-pion decays of the $ D^+ $ . From the Dalitz 
 plot analysis of the $D_s^+ \to \pi^- \pi^+ \pi^+$ decay 
 events, we find   significant contributions from the  
 channels $\rho^0(770)\pi^+$, $\rho^0(1450)\pi^+$, 
 $f_0(980)\pi^+$, $f_2(1270)\pi^+$,  and  
 $f_0(1370)\pi^+$. We also present  new measurement of 
 the masses and widths of the isoscalar resonances 
 $f_0(980)$ and $f_0(1370)$.

\vspace{1pc}
\end{abstract}


\maketitle

\section{Introduction}

E791 data were produced by 500 GeV/$c$ $ \pi^- $ interactions in five thin
foils (one platinum, four diamond) separated by gaps of 1.34 to 1.39~cm. 
The detector, the data set, the reconstruction, and the resulting
vertex resolutions have been described previously\cite{OLDTPL}.
After reconstruction, events with evidence of well-separated
production (primary) and decay  (secondary) vertices were retained for 
further analysis. From the 3-prong secondary vertex candidates,
we select a  $\pi^- \pi^+ \pi^+$ sample with invariant mass ranging 
from 1.7 to 2.1 GeV/c$^2$. For this analysis all charged particles are taken
to be pions; i.e., no direct use is made of particle identification.

\section{$D^+$ and  $D_s^+\to \pi^- \pi^+ \pi^+$ relative Branching Ratio}

\begin{figure}[htb]
\centerline{\epsfysize=2.00in \epsffile{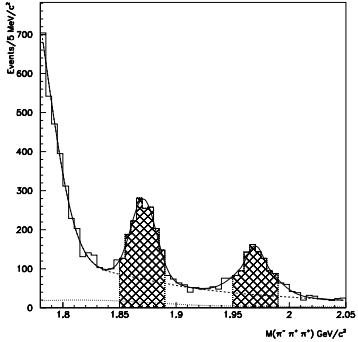}}
\caption{The $\pi^- \pi^+ \pi^+$ effective mass spectrum. The dotted line 
represents the $D^0 \to K^-\pi^+$ plus $ D_s^+ \to \eta^{\prime} \pi^+ $ 
and the dashed line is the total background. Events in the hatched areas
at the $D_s$ and $D^+$ mass are  used for the $D_s$ and $D^+$ Dalitz plot
analyses.}

\label{fig1}
\end{figure}

We fit the spectrum of Figure \ref{fig1} as the sum of $ D^+ $ and $ D_s^+ $ 
signals plus background. 
We model the background as the sum of four components: a general
combinatorial background, the reflection of the $D^+ \to K^-\pi^+\pi^+$ decay,
reflections of $D^0 \to K^-\pi^+$ plus one extra track (mostly from the primary
vertex),
and $D_s^+ \to \eta' \pi^+$  followed by
$\eta ' \to \rho^0(770) \gamma$, $\rho^0(770) \to \pi^+\pi^-$. 
We use Monte Carlo (MC) simulations to determine the shape 
  of each  identified  charm 
background in the $\pi^- \pi^+ \pi^+$ spectrum \cite{e791ds}.
We assume that the combinatorial background falls exponentially with mass.
 The fit  finds 1172  $\pm$  61  $D^+$ events and 
848 $\pm$  44 $D_s^+$ events. 

The branching fraction for  $D_s^+ \to \pi^- \pi^+ \pi^+$ relative to
        that for $D_s^+ \to \phi \pi^+$ is measured to be:

\begin{equation}
{ B(D_s^+ \to \pi^- \pi^+ \pi^+)  \over B(D_s^+ \to \phi \pi^+) } = 
 0.245 \pm 0.028^{+0.019}_{-0.012}\ .   
\end{equation}
\noindent

The branching fraction for $D^+ \to \pi^-\pi^+\pi^+$ 
relative to that of $D^+ \to K^-\pi^+\pi^+$ is measured to be:
\begin{equation}
{ B(D^+ \to \pi^- \pi^+ \pi^+)  \over{B(D^+ 
\to K^-\pi^+\pi^+)} } =  0.0311 \pm 0.0018 ^{+0.0016}_{-0.0026}\,. 
\end{equation}
The first error is statistical and the second is systematic.

\section { Dalitz plot formalism}

To study the resonant structure of these  decays  we consider the 
1686  events with invariant mass between 1.85 and 1.89 GeV,
for the $D^+ $ analysis and the 937 events with invariant mass
between 1.95 and 1.99 GeV/c$^2$ for the $D^+_s $.  Fig. 2a,b 
shows the Dalitz plot for these events. The horizontal and vertical 
axes are the squares of the $ \pi^+ \pi^- $ invariant masses, and 
the plot has been symmetrized with respect to the two $ \pi^+ $'s.

\begin{figure}[htb]
\centerline{\epsfysize=2.00in \epsffile{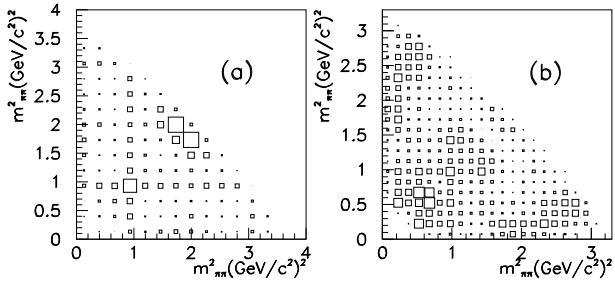}}
\caption{(a) The $D_s^+ \to \pi^- \pi^+ \pi^+$ Dalitz plot and (b) 
the $D^+ \to \pi^- \pi^+ \pi^+$ Dalitz plot. Since there are two
identical particles, the plots are symmetrized.}
\label{fig2}
\end{figure} 

We fit each distribution shown in Figure \ref{fig2} to a signal 
probability distribution function (PDF), which
is a coherent sum of  amplitudes corresponding to the non-resonant decay 
plus five different resonant channels, and a background 
PDF of known shape and magnitude. The resonant channels we include
in the fit are $\rho^0(770) \pi^+$, $f_0(980) \pi^+$, $f_2(1270) \pi^+$, 
$f_0(1370) \pi^+$, and   $\rho^0(1450) \pi^+$. 

We assume the non-resonant amplitude to be uniform across the Dalitz plot.
Each resonant amplitude, except that  for the $f_0(980)$, is
parameterized as a product of form factors, a relativistic Breit-Wigner 
function, and an angular momentum amplitude which depends on the spin 
of the resonance \cite{e791dp}. 

For the $f_0(980) \pi^+$ we use a coupled-channel Breit-Wigner function, 
following the parameterization of the WA76 Collaboration\cite{wa76},

\begin{equation}
BW_{f_0(980)} = {1 \over {m_{\pi\pi}^2 - m^2_0 + im_0(\Gamma_{\pi}+\Gamma_K)}}\ ,
\end{equation}
with 
\begin{equation}
\Gamma_{\pi} = g_{\pi}\sqrt{m_{\pi\pi}^2/ 4 - m_{\pi}^2}
\end{equation}
and
\begin{equation}
\Gamma_K = {g_K \over 2}\ \left( \sqrt{m_{\pi\pi}^2/ 4 - m_{K^+}^2}+
\sqrt{m_{\pi\pi}^2/ 4 - m_{K^0}^2}\right)\ .
\end{equation}

We multiply each amplitude  by a complex coefficient, $c_j=a_je^{\delta_j}$.
The fit parameters are the magnitudes, $a_j$, and the phases,
$\delta_j$, which accommodate the final state interactions. The reported 

 parameters values are  obtained using the maximum-likelihood method.

\section { $D_s^+ \to \pi^- \pi^+ \pi^+$ Dalitz plot  results}
The parameters of the $f_0(980)$ state, $g_{\pi}$, $g_K$, and $m_0$, as well 
as the mass and width of the $f_0(1370)$, are determined directly from the 
$D_s^+ \to \pi^- \pi^+ \pi^+$ decay events, floating them as free parameters 
in the fit. The measured values $m_0 =  977 \pm 3 \pm 2$ MeV/c$^2$, 
$g_{\pi} =$ 0.09  $\pm$  0.01  $\pm$  0.01 and  $g_K =$ 0.02  $\pm$  
0.04  $\pm$  0.03. For the $f_0(1370)$ we find 
$m_0 = $$1434 \pm 18 \pm 9$ MeV/c$^2$
and $\Gamma_0 =  173 \pm 32 \pm 6$ MeV/c$^2$. In both results, the first 
reported error is statistical and the second is systematic. The 
other resonance  masses and widths are taken 
from the PDG\cite{pdg}.

\begin{table}[htb]
\footnotesize{
 \begin{tabular}{|c|c|c|c|}     \hline
 mode &  relative phase  &  fraction(\%)
 \\ \hline \hline
   $f_0(980)\pi^+$& 0$^{\circ}$(fixed) &57. $\pm$ 4.  $\pm$ 5

 \\ \hline
   non-resonant  &(181. $\pm$ 94. $\pm$ 51.)$^{\circ}$ 
   &.5  $\pm$ 1.  $\pm$ 2.
 \\ \hline
  $\rho^0(770)\pi^+$ & (109. $\pm$ 24 $\pm$ 5.)$^{\circ}$  & 6.  $\pm$ 2.  $\pm$
  4
 \\ \hline
   $f_2(1270)\pi^+$  &(133. $\pm$ 13. $\pm$ 28.)$^{\circ}$&20.  $\pm$ 3.  $\pm$
   1.
 \\ \hline
   $f_0(1370)\pi^+$   &(198. $\pm$ 19. $\pm$ 27.)$^{\circ}$ &32.  $\pm$ 8  $\pm$
   2.
 \\ \hline
  $\rho^0(1450)\pi^+$  &(162. $\pm$ 26. $\pm$ 17.)$^{\circ}$ &4.  $\pm$ 2.  $\pm$ .2
 \\ \hline
\end{tabular}
\caption{Results of the $D^+_s$ Dalitz plot fits
(systematic error follows the statistical). }
}
\end{table}

 Table 1 shows  the phases
($\delta_j$) determined from the fit, and corresponding fraction for 
each decay mode. We calculate the decay fraction for each amplitude as
its intensity, integrated over the Dalitz plot, divided
by the integrated intensity of the signal's coherently summed amplitudes.

\begin{figure}[htb]
\centerline{\epsfysize=2.0in \epsffile{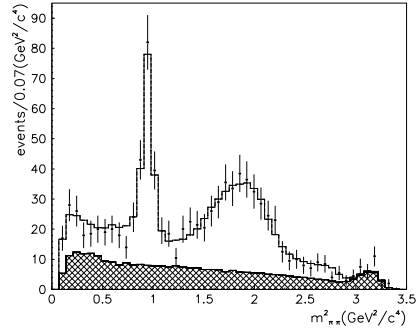}}
\caption{ $m^2_{\pi\pi}$  projections for data (dots) and fast-MC (solid).
The hashed area is the background distribution.}
\label{fig3}
\end{figure}

\noindent  The first 
reported error is statistical and the second is systematic, the latter being
dominated by the uncertainties in the resonance parameters, in the background
parameterization, and in the acceptance correction. 
The $f_0(980)\pi^+$ is the dominant component, accounting for nearly 
half of the $D_s^+ \to \pi^- \pi^+ \pi^+$ decay width, followed by
the $f_0(1370)\pi^+$ and $f_2(1270)\pi^+$ components. 
The contribution of  $\rho^0(770)\pi^+$ and $\rho^0(1450)\pi^+$ components 
corresponds to about 10\% of the $\pi^- \pi^+ \pi^+$ width. 
We have not found a statistically significant non-resonant component. 
The two $m^2_{\pi^+\pi^-}$  projections are nearly independent 
and the sum of them is shown in Fig. \ref{fig3}.

To assess the quality of our fit absolutely, and to compare it with other 
possible fits, we developed a fast-MC algorithm that simulates the 
$D_s^+ \to \pi^- \pi^+ \pi^+$ Dalitz plot from a given signal distribution,
background, detector resolution and acceptance.
For any given set of input parameters we calculated a $ \chi^2 $ using the
procedure presented in Ref. \cite{e791dp}.  
From $ \chi^2 $  and the number of degrees of freedom ($ \nu $),
we calculate a confidence level assuming a Gaussian distribution
in $ \chi^2 / \nu $.
The confidence level for the agreement of the projection of our result 
 onto the Dalitz plot with the data  of table I is 35\%.

\section { $D^+ \to \pi^- \pi^+ \pi^+$ Dalitz plot analysis results}

In a first model for this channel,
which we will refer to as Fit 1, the signal PDF includes 
the same resonances states used in the $D^+_s \to \pi^- \pi^+ \pi^+$ analysis. 
In the case of the $f_0(980)$ and $f_0(1370)$, we 
 used the parameters obtained in the $D^+ \to \pi^- \pi^+ \pi^+$
 analysis. In this model, the non-resonant, 
the $\rho^0(1450) \pi^+ $ and the $\rho^0(770) \pi^+ $ amplitudes  
 dominate \cite{e791dp}. The qualitative features of this fit are similar to
those reported by E691\cite{e691} and E687\cite{e687}.

\begin{figure}[t]
\centerline{\epsfysize=3.8in \epsffile{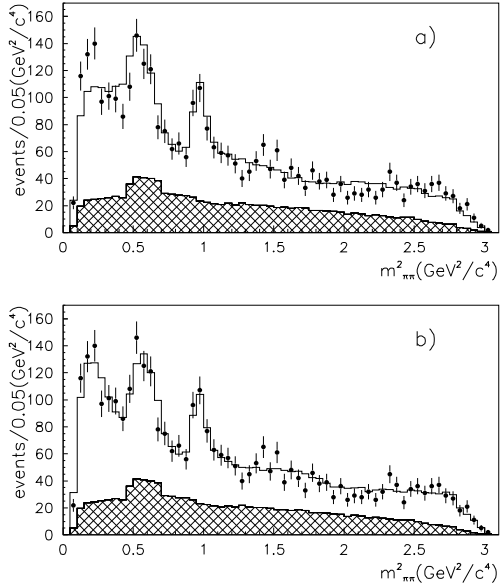}}
\caption{ $m^2_{\pi\pi}$  projection for data (error bars) and 
fast-MC (solid). The shaded area is 
the background distribution, (a) solution with the Fit 1, (b)  
solution with Fit 2.} 
\label{fig4}
\end{figure}

We produce the $ \chi^2 $ distribution for the difference 
in densities and observe a concentration of   $ \chi^2 $
in the low $ \pi^+ \pi^- $ mass$^2$ ($m^2_{\pi^+ \pi^-}$) region. 
The $ \chi^2 $ summed over all 
bins is 254 for 162 degrees of freedom ($ \nu $), which corresponds to
a confidence level less than $10^{-5}$, assuming Gaussian errors. 
 We display the sum of the two $m^2_{\pi\pi}$  in  Fig. \ref{fig4}a 
 for the data and for the fast-MC. 

The low confidence level casts doubt on the
validity of the model used.
While the projection of the  MC onto the $ \pi^+ \pi^- $
mass${}^2$ axis describes the data in the $ \rho^0 (770) $ and
$ f_0(980) $ regions well, there is a discrepancy at lower mass,
suggesting the possibility of another  amplitude.
To investigate the possibility that another 
$ \pi^+ \pi^- $ resonance contributes
an amplitude to the $ D^+ \to \pi^- \pi^+ \pi^+ $ decay,
we add a sixth resonant amplitude to the signal PDF.
We allow its mass and width to float and assume
a scalar angular distribution.
This fit (Fit 2) converges and finds values 
of $ 478^{+24}_{-23} \pm 17$  MeV$/c^2$ for the mass
and  $ 324^{+42}_{-40} \pm 21$ MeV$/c^2$ for the width,
the first error is statistical and the second systematic.
We will refer to this possible state as the $ \sigma(500) $.

In Fit 2, the $ \sigma (500) $ amplitude produces the largest
decay fraction, 46\%, with a relatively small statistical error,
9\%. The non-resonant fraction, which at $ (39 \pm 10) \% $ was
the largest in the original fit, is now only $ (8 \pm 6 ) \% $.
When we project this model onto the Dalitz plot, the
$ \chi^2 / \nu $ becomes 138/162.
The projection of this model onto the $ \pi^+ \pi^- $ invariant mass
squared distribution, shown in Fig. \ref{fig4}b,
describes the data well, including the accumulation of events near
0.2 GeV$^2/c^4$.

\begin{table}[htb] 
\footnotesize{
 \centering
 \begin{tabular}{|c|c|c|c|}     \hline
 mode  & relative phase  &  fraction(\%)
 \\ \hline \hline
  $\sigma\pi^+$ & (207. $\pm$ 8. $\pm$ 5.)$^{\circ}$ 
  & 46. $\pm$ 9. $\pm$ 2.
 \\ \hline
   $\rho^0(770)\pi^+$ &0$^{\circ}$(fixed) &34. $\pm$ 3. $\pm$ 2.

 \\ \hline
   non-resonant  & (57. $\pm$ 20. $\pm$ 6.)$^{\circ}$
   &8. $\pm$ 6. $\pm$ 3.
 \\ \hline
  $f_0(980)\pi^+$   &(165. $\pm$ 11. $\pm$ 3.)$^{\circ}$
  &6. $\pm$ 1. $\pm$ .4
 \\ \hline
   $f_2(1270)\pi^+$   &(57. $\pm$ 8. $\pm$ 3.)$^{\circ}$
   & 19. $\pm$ 3. $\pm$ 0.4
 \\ \hline
   $f_0(1370)\pi^+$   &(105. $\pm$ 18. $\pm$ 1.)$^{\circ}$
   &2.5 $\pm$ 1.5 $\pm$ 1.
 \\ \hline
  $\rho^0(1450)\pi^+$  &(319. $\pm$ 29. $\pm$ 11.)$^{\circ}$
  &1. $\pm$ 1. $\pm$ .3
 \\ \hline
\end{tabular}
\caption{Results of the $D^+$ Dalitz plot fit,
including the scalar $\sigma(500)$ resonance.
(the systematic error follows the statistical). }
}
\end{table}

To better understand our data,
we also fit it with vector, tensor, and toy models
for the sixth (sigma) amplitude, allowing the masses, widths, and 
relative amplitudes to
float freely.
The vector and tensor models test the angular distribution of the
signal.
The toy model tests the phase variation expected of a Breit-Wigner amplitude
by substituting a constant relative phase. These alternative explanations 
of the data fail to describe it as well the scalar with a regular Breit 
Wigner option \cite{e791dp}.

\section { Conclusions}

 Using a Dalitz plot analysis  of the three pion decay of the 
 $D_s^+$, we find a little but  significant contributions 
 from the  channels $\rho^0(770)\pi^+$, $\rho^0(1450)\pi^+$. We  measure 
 the mass and width of the $f_0(980)$ 
 and we have not found a statistically significant $g_K$ component 
 in the width. Using the same  analysis for the $D^+$ mass region,
we find strong evidence that a scalar resonance with mass
$478^{+24}_{-23} \pm 17 $ MeV/$c^2$ and width
$324^{+42}_{-40} \pm 21 $ MeV/$c^2$
produces a decay fraction $ \approx 50\% $. These results are 
discussed  in references \cite{e791ds,e791dp,dib}.

\small
\bibliographystyle{unsrt}
 
\end{document}